\documentclass[twocolumn,10pt]{article}
\usepackage[a4paper,margin=1in]{geometry} % Set margins
\usepackage{amsmath, amssymb, amsthm} % Math packages
\usepackage{graphicx} % Graphics
\usepackage{authblk} % Author and affiliation
\usepackage{hyperref} % Hyperlinks
\usepackage{abstract} % Abstract formatting
\usepackage{enumitem} % Custom lists
\usepackage{titlesec} % Section title customization
\usepackage{color} % Color for text highlighting if needed
\usepackage{setspace} % Line spacing for better readability
\usepackage[square,numbers]{natbib} % For references with numbers
\usepackage{soul} % For highlighting

\newcommand{\phflo}[1]{\textcolor{black}{#1}}
\newcommand{\mflo}[1]{\textcolor{black}{#1}}

\title{Diceplot: A package for high dimensional categorical data visualization}
\author[1,2]{Matthias Flotho}
\author[1]{Philipp Flotho}
\author[1,2]{Andreas Keller}
\affil[1]{Chair for Clinical Bioinformatics, Center for Bioinformatics, Saarland University, Saarland, Germany}
\affil[2]{Helmholtz Institute for Pharmaceutical Research Saarland (HIPS), Saarland University Campus, Saarland, Germany}
\date{}

\begin{document}

\twocolumn[
    \maketitle
\begin{abstract}{
Visualization of multidimensional, categorical data  is a common challenge \mflo{across} scientific areas and, in particular, the life sciences. The goal is to create a comprehensive overview of the underlying data which allows to assess multiple variables intuitively.
%This is not limited to but holds in particular for pathway analysis, checking for dysregulation in known biological regulatory mechanisms and functions, across multiple conditions.
\phflo{One application where such visualizations are particularly useful is pathway analysis, where we
}
check for dysregulation in known biological regulatory mechanisms and functions across multiple conditions.
Here, we propose a new visualization approach that codes such data in a comprehensive  and intuitive representation: \mflo{Dice plots visualize up to four distinct categorical classes in a single view that consist of multiple elements resembling the faces of dice, whereas domino plots add an additional layer of information for binary comparison.}
%The code is available in python as well as R and is available at \url{https://github.com/maflot/diceplot} and \url{https://github.com/maflot/pydiceplot}.
\phflo{The code is available as the \textbf{diceplot} R package, as \textbf{pydiceplot} on pip and at \url{https://github.com/maflot}.}
}
\textbf{data visualization, pathway analysis, python, R, scRNA-seq}
\end{abstract}
\vspace{1em}
]

\section{Introduction}
In bio-sciences, visualizing and representing data in an appealing and informative way is crucial for %exploring and presenting the data. 
\phflo{data exploration and presentation.}
\mflo{In single-cell RNA-sequencing studies, we face the problem that we examine different conditions and \phflo{require} visualizations that allow the simultaneous assessment of multiple variables.}
In particular, visualizing pathway analysis's across multiple cell types and conditions is challenging as we are often interested not only in the intersections across diverse conditions but also in the categories \phflo{of each variable}.
\mflo{There exist \phflo{many established and} excellent solutions for high-level data visualization. \phflo{An important example are} scatter plots of UMAP or PCA embeddings\phflo{, which are used to visualize the structure} of the underlying \phflo{high-dimensional} data. Venn diagrams \cite{venn1880diagrammatic} \phflo{are useful to} highlight the intersections of different conditions. Moreover, approaches such as upset plots \cite{lex2014upset} are brilliant for visualizing the quantitative overlaps and present a great improvement for handling many sets to the classical Venn diagram for diverse groupings. Scalability and selection of the intersecting sections is not trivial. \phflo{Hence,} more dynamic approaches are required \cite{amand2019dynavenn}. Finally, another well established plot is the circle plot, which offers great options for visualizing and quantifying intersections between several groups \cite{gu2014circlize}. \phflo{While all of those approaches are great to visualize the quantitative overlap of groups, they lose any} information about the elements of the intersections.}
\mflo{\phflo{For qualitative analyses, there exist} plenty of plots and solutions to highlight and display a detailed look on the data and examined conditions, and or cell types, or the expression of single genes.
\phflo{Important examples include} bar, or scatter plots. For example, volcano plots are brilliant to visualize and highlight significantly dysregulated genes between two conditions, or simple scatter plots of the gene expression of two genes for a simple correlation analysis.
\phflo{However, those approaches can again only visualize one aspect of the data.}
\phflo{They are useful to visualize qualitative details of the data but are not useful for high-level overviews} of the data. Here we present dice- and dominoplots, an intuitive data visualization aiming to bridge the high-level and the low-level view of the data. The representation allows to display up to four different dimensions of the data in a dice plot and double the amount in a domino plot. This will serve} as an addition to the established plotting approaches. This representations enables a comprehensive overview about the quantity of shared attributes while keeping the the set information visible. We will show on the example of pathway analysis how to use and interpret the dice and domino plots.

\section{Diceplot}\label{sec2}

\begin{figure*}[!t]%
\centering
    \includegraphics[width=1\linewidth]{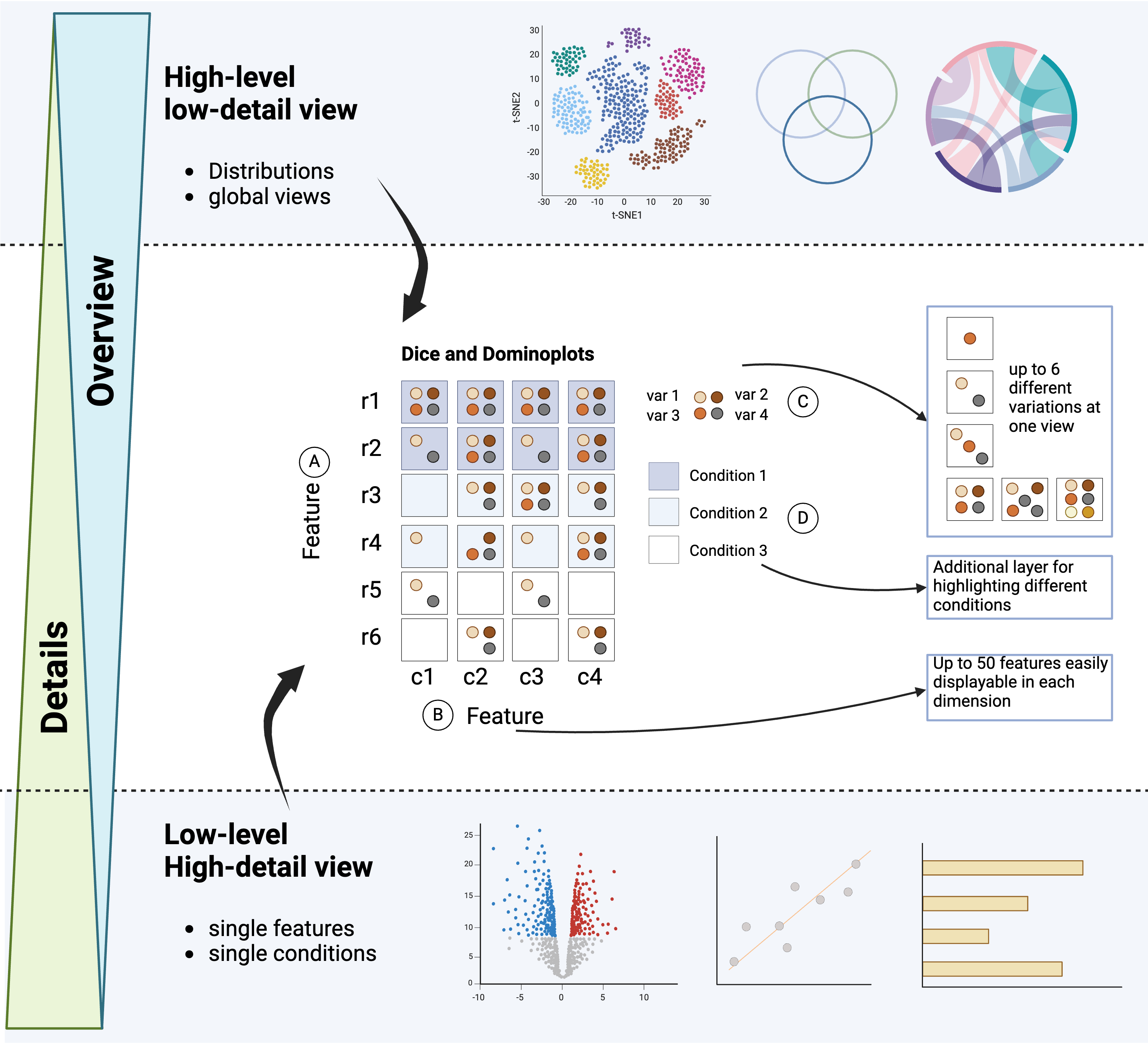}
    \caption{The diceplot package enables a valuable trade-off between a high-level view of the data displaying the broader structure of the underlying data and low-level views of the data highlighting single features or conditions. A dice plot allows to display up to four different categorical variables (A,B,C, and D).
    Here, the user can chose between different visualizations to order and group the features in A, and B. C can be used to display up to 6 distinct categories of a categorical variable. Additionally, the underlying boxes can be colored for encoding an additional variable (D).}
    \label{fig:diceplot}
\end{figure*}

Displaying highly dimensional categorical data is challenging without splitting the information into diverse subplots or losing categorical information. 
Here we are able to display up to four distinct categorical variables at a single view.
\mflo{For emphasizing this we will take a look at the example of pathway analysis across different pathways for feature (A), and cell types for feature (B). In this example the aim is to visualize the pathways for the different cell types, across different disease variants (C). Finally, the diceplot allows to keep a high-level grouping for the pathways (D) (Fig. \ref{fig:diceplot}). 
The objective is to visualize all this information in a single view.
Grouping the different variations based on the sites of a dice allows the user to visualize up to six different groups once.  The coloring of the different groups can be each either categorical or continuous, as the variation information is already encoded int the position of the dots.}
The dot arrangement is fixed which allows an easy overview of whether the groups are present or not. The coloring can be used to further enrich the plot with continuous color scales or categorical palettes.
Those dice can be colored with a separate background color to transport further information to the user, in our example this represents a higher grouping of the pathways examined.
This plot breaks down the information in a clear and comprehensive way.

\section{Dominoplot}\label{sec3}

To even further extend the information for a binary comparison and for comparing continuous variables. We extended the frame work for dominoplots.
Dominoplots allow an additional level added by placing two dice next to each other forming a domino. For example, plotting the gene dysregulation for two separate setups and multiple conditions such as the comparison of different conditions across sexes. Additional to the color coding, we implement varying dot sizes for highlighting continuous information.
Here, the user could use an additional layer to visualize their findings side by side without losing any further information. We recommend discarding the dice background coloring to not overcrowd the plot, but the user is free to chose to ignore this recommendation.

\section{Implementation}
\mflo{We maximize the usability of our approach by providing an R as well as a python implementation. Moreover the code will be easily available via the \textit{Comprehensive R Archive Network} (CRAN) \cite{r2013r} and the \textit{python package index} (pip) \cite{pypi}. Moreover, different back-ends enables an universal and easy use of the package. In particular, the plotly back-end for the pyDicePlot is easy to use and easily applicable for web interfaces and interactive exploration of the data.}

\section{Conclusion}

\mflo{This note presents an easy to use plot library to address the visualization of multiple categorical dimensions in one view. This effectively bridges the gap between high-level views, providing a broad overview of the data, and low-level views diving into the details of the data. As example use-case, our approach enables pathway analysis on a new level keeping large parts of the information visible which would be lost using common plots as upset-plots or chord diagrams. In those two examples we can beautifully display the fractions of shared pathways but lose inevitably the information of the actual pathways. Our approach significantly, improves the visualization of the data, still it is limited in its number of features for each of the dimensions. While in theory there is no limitation for the number of rows and columns to display, in practice consideration is needed to select reasonable number of rows and columns to keep the visualization clean and comprehensive. In summary, diceplots are a great addition to the existing of visualizations highlighting a mid-level view of the examined data. In future work, we plan to set up a interactive plotting suite combining other visualizations with the dice- and dominoplots accessible via an webserver. This will further make visualization more accessible for researchers not familiar with R and python programming.}

\section{Code and data availability}\label{sec6}

The code of the package is available at  \url{https://github.com/maflot/DicePlot} (\textbf{R} implementation) and
\url{https://github.com/maflot/pyDicePlot/} (\textbf{Python} implementation). The Python package is also available as pydiceplot package on pip.
Use-case scenarios and additional documentation can be found at \url{https://dice-and-domino-plot.readthedocs.io/en/latest/}

\section{Acknowledgements}\label{sec6}
The figure was created in  BioRender.com.
This study was financed through the DFG project 469073465.
%%%%%%%%%%%%%%

\section{Competing interests}
No competing interest is declared.

\vspace{-5pt}
\section{Author contributions statement}

M.F. implemented and invented the representation, P.F. supported the python implementation and contributed to the manuscript,  A.K helped conceptualizing, refining, and finalizing the development process.

%\bibliographystyle{plain}\part{title}
%\bibliography{reference}

%USE THE BELOW OPTIONS IN CASE YOU NEED AUTHOR YEAR FORMAT.
%\bibliographystyle{abbrvnat}
%\bibliography{reference}

\end{document}